\documentclass[a4paper,11pt]{article}
\usepackage{pos}
\newcommand{\amuhlo}{$a_{\mu}^\text{HLO}$}

\newcommand{\gmtwo}{\ensuremath{g\!-\!2}}
\newcommand{\amu}{$a_{\mu}$}
\newcommand{\omegaa}{$\omega_a$}
\newcommand{\omegap}{$\omega_p$}

\title{New results from the Muon \gmtwo\ Experiment}

\author[a,b,*]{Graziano Venanzoni}

\affiliation[a]{University of Liverpool, Liverpool L69 3BX, United Kingdom}

\affiliation[b]{INFN Sezione di Pisa, Largo Bruno Pontecorvo 3, 56127, Pisa, Italy}

\affiliation[*]{On behalf of the Muon \gmtwo\ Collaboration}

\emailAdd{graziano.venanzoni@liverpool.ac.uk}

\abstract{
The Muon \gmtwo\ experiment at Fermilab has published the first result on Run-1 dataset in 2021
showing a good agreement with the previous experimental result at Brookhaven National Laboratory at comparable precision (0.46 ppm). In August 2023 we released our new result from  Run-2 and Run-3 datasets which allowed to measure \amu\  to 0.21 ppm, a more than two-fold improved precision respect to Run-1, and which allowed to reach a precision of 0.20 ppm when combined with the Run-1 result.
We will discuss the improvements of the Run-2/3 analysis respect to Run-1, the current status of the theory prediction, and the future prospects.
}

\FullConference{The European Physical Society Conference on High Energy Physics (EPS-HEP2023)\\
 21-25 August 2023\\
Hamburg, Germany\\}

\begin{document}
\maketitle

\section{The Muon \gmtwo}
The muon anomalous magnetic moment – the so-called muon \gmtwo, where $g$ is the gyromagnetic ratio or the g-factor of the muon – is a particle physics observable where a significant discrepancy between measurement and the Standard Model (SM) persists for more than 20 years~\cite{Jegerlehner:2017gek}. 
The E821 collaboration at Brookhaven National Laboratory (USA) has determined the muon magnetic anomaly $a_\mu = (g-2)_\mu/2$ with a 
relative precision of 0.54 parts per million (ppm) and found a discrepancy with the SM prediction of less than 3 standard deviations~\cite{Muong-2:2006rrc}. This discrepancy has increased in the following years, as a result of improved accuracy of the SM prediction, reaching the level of 3.7 standard deviations~\cite{Aoyama:2020ynm}, representing one of the largest deviations between data and SM prediction among the whole set of experimental measurements in particle physics and a possible harbinger of physics beyond the SM~\cite{Czarnecki:2001pv,Stockinger:2006zn}. Given this discrepancy a new Muon \gmtwo\ experiment, E989, was built at Fermilab (USA), and started data taking in 2018 with the goal to reduce the experimental uncertainty by a factor of four~\cite{Muong-2:2015xgu}. In addition a new Muon \gmtwo\ experiment with a completely different approach is currently under construction at J-PARC (Japan)~\cite{Abe:2019thb}.

\noindent On the theoretical side, the evaluation of the SM prediction of \amu\ has occupied many physicists for over seventy years~\cite{Jegerlehner:2017gek}. 
Since Schwinger's famous calculation of the electron's g-factor in 1948~\cite{Schwinger:1948iu}, a heroic effort has been put into the theoretical prediction of \amu\ commensurate with the experimental accuracy. Today, the state of the art is the inclusion of the QED corrections up to five so-called loops (virtual exchanges with photons and fermions - Schwinger did one loop where just one virtual photon is exchanged) and the evaluation of the electroweak and hadronic contributions, which allowed to reach a precision of 0.37 ppm~\cite{Aoyama:2020ynm}. Recent evaluations and new results of the leading-order hadronic contribution to the muon \gmtwo\ have challenged the evaluation~\cite{Aoyama:2020ynm} (see Section~\ref{theory}).

\section{The Muon \gmtwo\ experiment at Fermilab}
\begin{figure}[!htb]
    \centering
    \includegraphics[width=.5\textwidth]{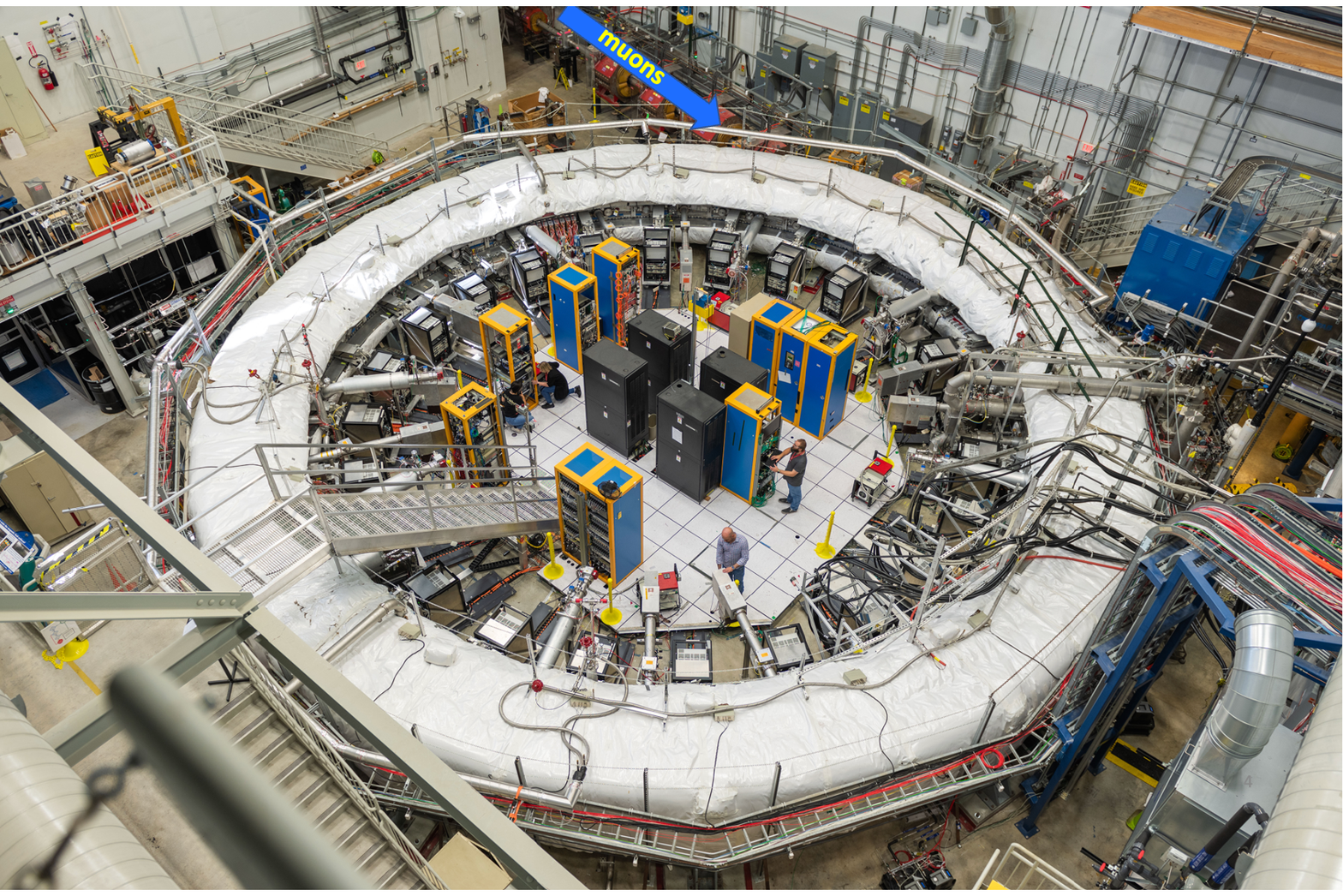}
    \caption{Aerial view of the Muon \gmtwo\ experimental hall showing the muon injection and the storage ring covered with the thermal insulating blanket.}
    \label{ring}
\end{figure}

The Muon \gmtwo\ experiment uses polarized muons produced at Fermilab's Muon Campus~\cite{Stratakis:2019sun}.  A beam of $8$ GeV protons is focused onto an Inconel target.  Positive pions with a momentum of $3.1$ GeV$/c$ coming off target are selected and directed along a nearly 2000 m decay line, including several revolutions around the Delivery Ring, which are used to further eliminate pions and to displace secondary protons from muons using time of flight and a kicker to sweep out the protons.  The resulting pure muon beam with 95\% polarization is injected into the storage ring  at an average rate of about $12$ Hz.  The muon storage ring is recycled from the Brookhaven E821 experiment~\cite{Danby:2001eh}.  The magnetic field is generated by three superconducting coils that follow the circumference of the steel yoke leading to a continuous field around the ring.  
Figure~\ref{ring} shows a top view of the Muon \gmtwo\ experimental hall.
Muons enter the ring through a superconducting inflector magnet and are placed in the orbit by a set of three fast magnetic kickers.
Four Electrostatic Quadrupoles (ESQ) symmetrically placed around the ring each with a long and short arm provide vertical focusing.
24 Fast electromagnetic calorimeters made of lead fluoride (PbF$_2$) crystals, with large area (1.2 x 1.2 cm$^2$) Silicon Photo-Multiplier (SiPM) readout, placed inside the ring detect inward-spiraling decay positrons.  A state-of-the-art laser calibration system provides a control of the gain fluctuations. Two straw tube trackers to precisely monitor properties of stored muons are placed in the vacuum chamber at 180 and 270 degrees respect to the injection point.
The $\pm$ 25 ppm uniform magnetic field coupled with the ESQ acts as a 15-meter Penning trap: in the time window of $700 \mu s$ muons circulate for hundreds to thousands of turns. During the motion muons behave as spinning tops and we measure two quantities: the anomalous precession frequency which is the rate at which the spin of the muon advances respect to the momentum, and the magnetic field. The measurement of the precession frequency is made by counting the positrons with an energy above a certain energy threshold as a function of time. Due to the parity violation in the weak process of muon decay, high-energy positrons are emitted preferably towards the muon’s spin direction which makes the counting rate oscillate with \omegaa\ frequency, with its maximum occuring when the muon spin and momentum vectors are aligned. \omegaa\  can therefore be extracted by fitting the histogram of the
oscillating number of counted positrons as a function of time, also called “wiggle plot”, with a multi-parameter function that takes into account, amongst others, many beam dynamics effects. 
The measurement of the magnetic field is based on nuclear magnetic resonance (NMR) probes: 378
fixed probes are distributed in many azimuthal positions of the storage ring and placed on the
walls of the vacuum chambers, to continuously track field drifts. A set of 17 probes are inserted in 
a moving device, the so-called “field trolley”, that is pulled through the ring every 3 -- 5 days
without the muon beam, to measure the spatial field distribution in the storage region. The magnetic field is expressed in terms of the precession frequency (Larmor frequency) \omegap\ of protons at rest shielded in water sample. 

\noindent From the ratio of the two frequencies \omegaa\ and \omegap\ we extract \amu:

\begin{equation}\label{eq:e989_a_mu}
    a_{\mu} = \frac{\omega_a}{\omega_p}\,\frac{\mu_p}{\mu_e}\,\frac{m_{\mu}}{m_e}\,\frac{g_e}{2}
\end{equation}

\noindent
The first ratio $\omega_a/\omega_p$ is obtained directly in our experiment, measuring $\omega_a$ through the wiggle plot fit and $\omega_p$ with nuclear magnetic resonance (NMR) probes. The other quantities are well-known constants from other experiments, which carry a small overall uncertainty of $\sim25\,$parts per billion (ppb). Beam dynamics effects arising by the real motion of the muons  and the presence of transient magnetic fields make the formula more complex~\cite{Muong-2:2023cdq}:

\begin{equation}\label{eq:master_formula}
    a_{\mu} = \underbrace{\left[\frac{f_{clock}\cdot\omega^m_a\left(1+C_e+C_p+C_{pa}+C_{dd}+C_{ml}\right)}
    {f_{calib}\cdot\langle\omega_p\left(\vec{r}\right)\times M\left(\vec{r}\right)\rangle\left(1+B_q+B_k\right)}\right]}_{\mathcal{R_{\mu}}}\,\frac{\mu_p(T_r)}{\mu_e(H)}\,\frac{\mu_e(H)}{\mu_e}\,\frac{m_{\mu}}{m_e}\,\frac{g_e}{2}
\end{equation}

\noindent At the numerator, $\omega^m_a$ is the measured value of the precession frequency from wiggle plot fits, and it is multiplied by correction factors $C_i$ that come from beam dynamics. At the denominator, $\omega_p$ is weighted by the muon beam spatial distributions, and corrected by two fast magnetic transients $B_i$, from kickers and quadrupoles, synchronized to the beam injection. The unblinding factor $f_{clock}$ is set and monitored by external and unknown people to the Muon \gmtwo\ collaboration, and it is in the range $\pm\,25\,$ppm, to prevent possible biases on our measurement. $T_r = 34.7^\circ$C is the reference temperature at which the shielded proton-to-electron magnetic moment $\mu_p(T_r)/\mu_e(H)$ is measured~\cite{shielded_proton_electron}. The QED factor $\mu_e(H)/\mu_e$ is the ratio of the magnetic momentum of the electron in a hydrogen atom to the magnetic momentum of the free electron in vacuum~\cite{Karshenboim_codata,Tiesinga:2021myr}. The ratio in masses $m_{\mu}/m_e$ was measured to $22\,$ppb with muonium spectroscopy~\cite{muon_mass_ratio_codata,Tiesinga:2021myr}, and new precise measurements from several experiments (such as MuSEUM at J-PARC~\cite{MuSEUM} and Mu-MASS at PSI~\cite{mu-mass}) are expected soon. $\mathcal{R_{\mu}}$ is what we measure in our experiment.

\section{Data campaigns}\label{sec:run_conditions}

\begin{figure}[htb]
    \centering
    \includegraphics[width=.8\textwidth]{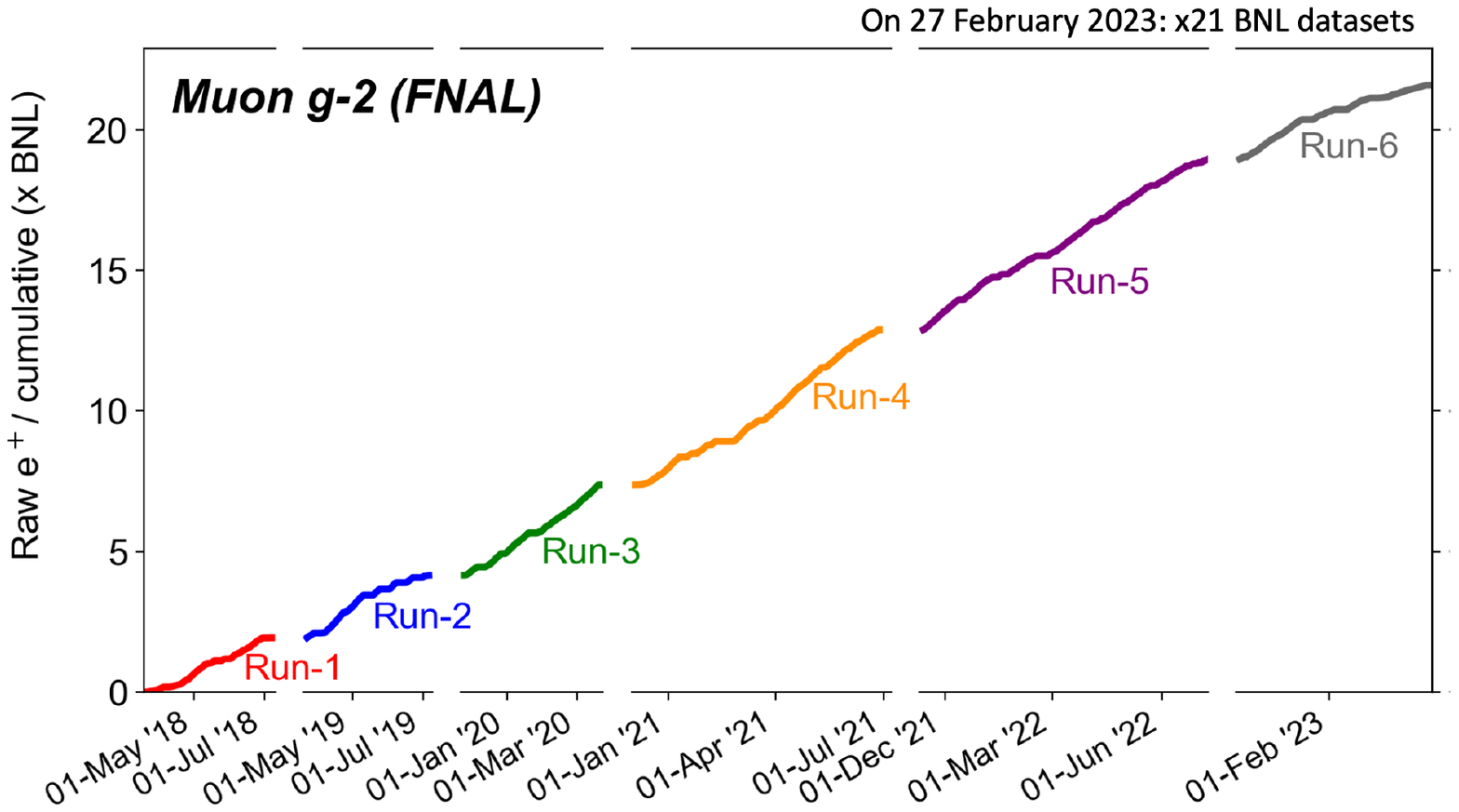}
    \caption{The $6$ run periods of the E989 experiment, with labels indicating the periods in which each data acquisition campaign took place. The last day of acquisition was $9^{th}$ July 2023 while on the $27^{th}$ February 2023 the statistical goal of 21$\times$ BNL dataset was reached.}
    \label{dataset}
\end{figure}

In the $6$ years of running the Muon \gmtwo\ experiment at Fermilab collected $21.9$ times the number of positrons than the previous experiment E821 at Brookhaven which would allow us to reach and possibly surpass the design goal of $100\,$ppb for the statistical uncertainty on \amu~\cite{Muong-2:2015xgu}. Figure \ref{dataset} shows how our data is split into $6$ campaigns, from 2018 to 2023. Each dataset is characterized by different running conditions such as ESQ voltage, kicker settings, muon beam dynamics, muon storing efficiency. The last data campaign, Run-6, was completed on the $9^{th}$ July 2023.

\section{Run-1 result}
In April 2021 the Muon \gmtwo\ experiment released its first result, based on the statistics collected in Run-1 (from April to June 2018) roughly corresponding to 1 BNL. Our result~\cite{Muong-2:2021ojo} confirmed the predecessor E821 BNL result and showed a more than 4 standard deviations difference respect to the SM prediction compiled in 2020~\cite{Aoyama:2020ynm}, which at that time~\footnote{The situation on the theory has changed since our first result, as it will be discussed in Section~\ref{theory}.} was a significant indication of possible New Physics.

\section{Improvements of Run-2/3 respect to Run-1 analysis}

Run-2 and Run-3 data campaigns were collected respectively from March to July 2019 and from November 2019 to March 2020~\footnote{Run-3 was planned to last until July 2020, but ended before due to the COVID-19 pandemic.}. 
Run-2/3 were divided into a total of $20$ datasets, but, thanks to the improved stability of the hardware conditions with respect to Run-1, many datasets were combined to allow higher statistics in the $\omega_a$ and reduction of the statistical uncertainties of some systematic effects. Thus, Run-2/3 were divided into three major datasets for analysis: Run-2, Run-3a and Run-3b. On the B-field side, a total of $25$ and $44$ trolley runs have been performed for Run-2 and Run-3, respectively, as opposed to $19$ trolley runs in Run-1.\\
Due to more than 4 times the number of positrons resulting from muon decay compared to the Run-1 result, Run-2/3 reduces the statistical uncertainty to 0.20 ppm, as shown in Fig~\ref{run23}, Left. Combined with Run-1, Run-2/3 statistics allows to achieve a statistical uncertainty of 0.18 ppm.

\noindent We were able to reduce the systematic error by more than a factor of two mainly due to better running conditions, dedicated systematic studies, and improvements on the analysis:

\begin{figure}[htb]
    \includegraphics[width=.55\textwidth]{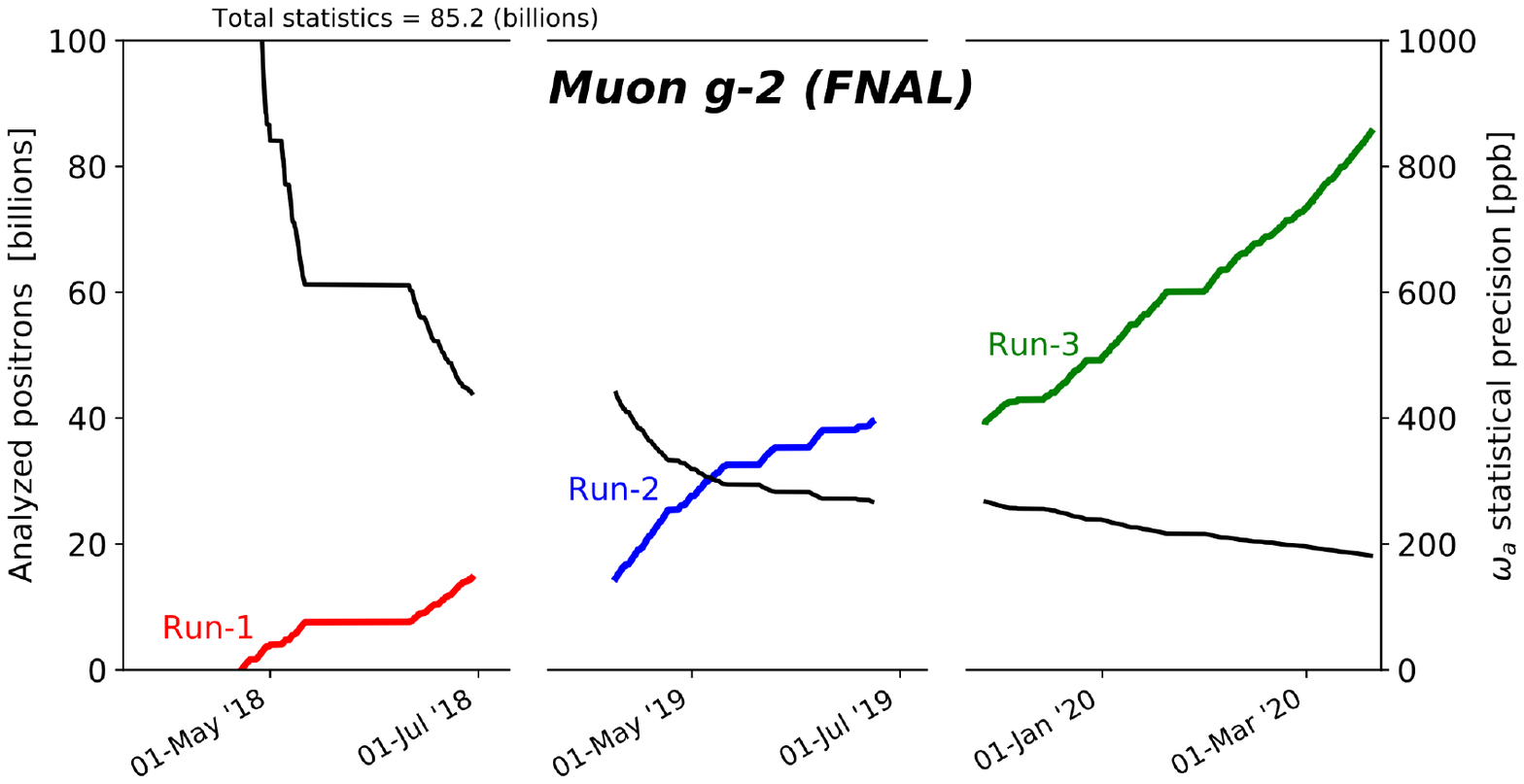}
    \includegraphics[width=.55\textwidth]{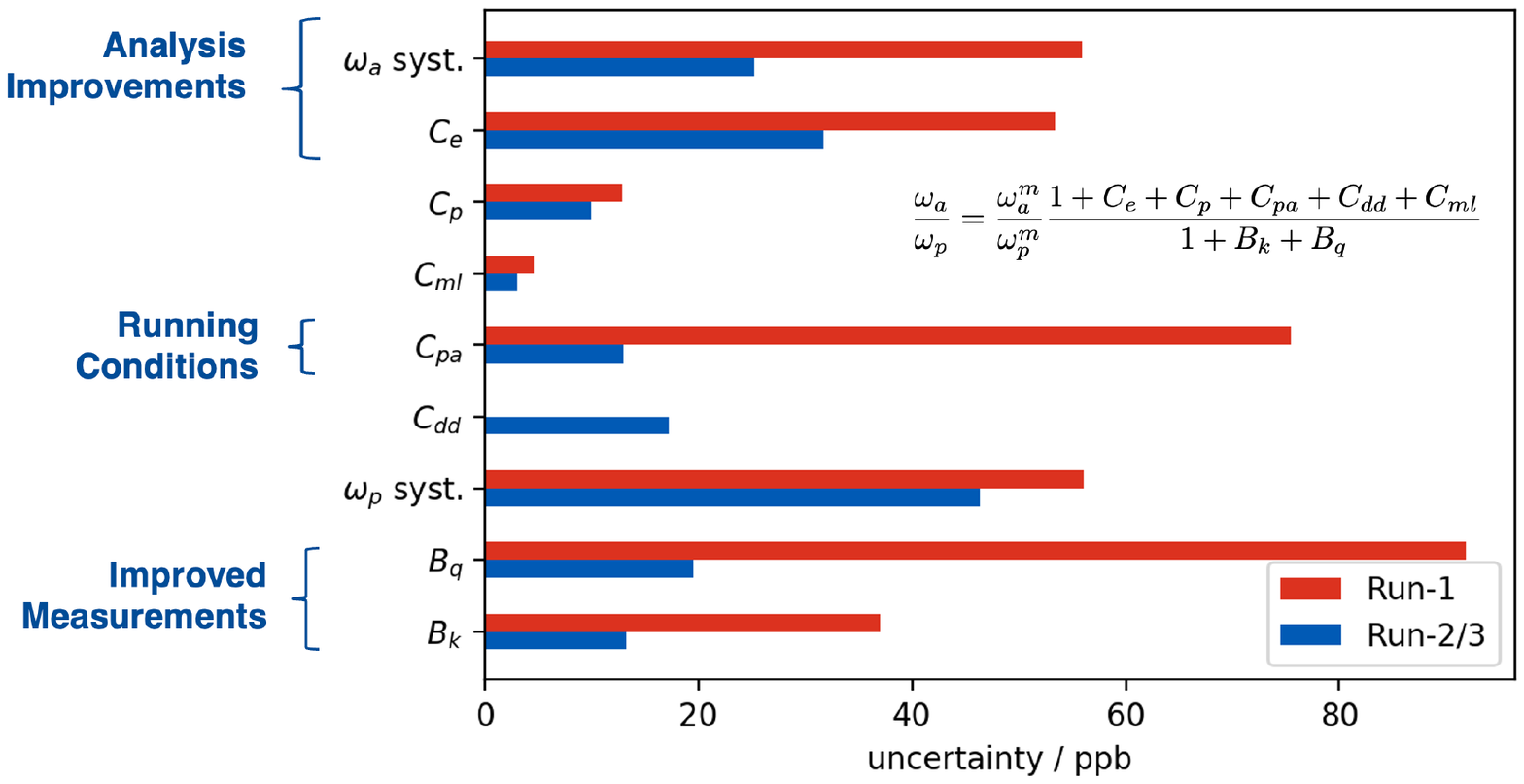}  
    \caption{Left: Run-2/3 analysed positrons from muon decays and statistical precision. Right: Run-2/3 systematic errors compared to Run-1.}
    \label{run23}
\end{figure}

\begin{itemize}
   \item Improvements on the running conditions:
        \begin{itemize}
            \item Two damaged resistors in the ESQ plates were replaced at the end of Run-1, improving the stability of radial and vertical betatron oscillations and reducing the phase acceptance correction $C_{pa}$ in Run-2/3 significantly;
            \item The kicker strengths for Run-1 and Run-2 were limited to {142} {kV} by the cables used at that time: as a result, the beam was not perfectly centered in the storage region. At the end of Run-3a, the cables were upgraded and the kicker voltage was increased to {165} {kV} in Run-3b to achieve optimal kick. This resulted in a better-centered muon beam which reduced the electric field $C_e$ correction;
            \item Between Run-1 and Run-2 the magnet yokes have been covered with a thermal insulating blanket to mitigate day-night field oscillations due to temperature drifts. In addition, the experimental hall's air conditioning system has been upgraded after Run-2 to further stabilize the temperature of both the magnet yokes and the detector electronics to better than $\pm{0.5} ^\circ$C;
            \item The number of lost muons was greatly reduced in Run-2/3 thanks to two upgrades. Firstly, the operational high-voltage set points for the ESQ system were lowered, in order to avoid betatron resonances for beam stability. Secondly, all $5$ collimators were used in Run-2/3, whereas only $2$ were used in Run-1: this allowed for better beam scraping;
        \end{itemize}
    \item Improved systematic studies:
        \begin{itemize}           
            \item  The correction from the magnetic field transient due to vibrations caused by ESQ pulsing, $B_q$, was
            measured at a much larger azimuthal locations around the ring. This mapping, in combination with improved methodology and repeated measurements over time, reduced  $B_q$ systematic uncertainty by more than a factor of 4 with respect to Run-1;
            \item An improved magnetometer, with a better setup and reduced vibration noise, reduced the systematic error on $B_k$, caused by kicker-induced eddy currents, by a factor of $\sim 3$;       
        \end{itemize}
    \item Analysis improvements:
        \begin{itemize}
            \item    The largest reduction on \omegaa\ analysis comes from the treatment of pileup, when two positrons enter a calorimeter close in time and are not separated by reconstruction algorithms.  Improved clustering of crystal hits in the reconstruction algorithms coupled with improved analysis methods reduced the number of unresolved pileup events and allowed to reduce the pileup uncertainty from 35 ppb in Run-1 to 7 ppb in Run-2/3;  
            \item  The largest beam dynamics correction, $C_e$, due to the electric fields of the ESQ system,  depends on the momentum spread of the muon beam. The muon momentum distribution is determined from the frequency distribution and debunching rate of the injected beam using calorimeter data, and the radial distribution of stored muons over a betatron period is obtained from tracker data. In Run-2/3, the debunching analysis took into account differences in momentum spread along the injected bunch length that were not included in the Run-1 analysis. Accounting for this difference and using complementary tracker information reduced the $C_e$ uncertainty by a factor of $\sim 1.6$
        \end{itemize}
\end{itemize}

\noindent As shown in Fig.~\ref{run23}, Right, all the sources of systematic error were significantly reduced for a combined systematic uncertainty of 70 ppb, which surpassed  our proposal goal of 100 ppb~\cite{Muong-2:2015xgu}.

\section{Run-2/3 result}
\begin{figure}[htb]
   \centering
    \includegraphics[width=.55\textwidth]{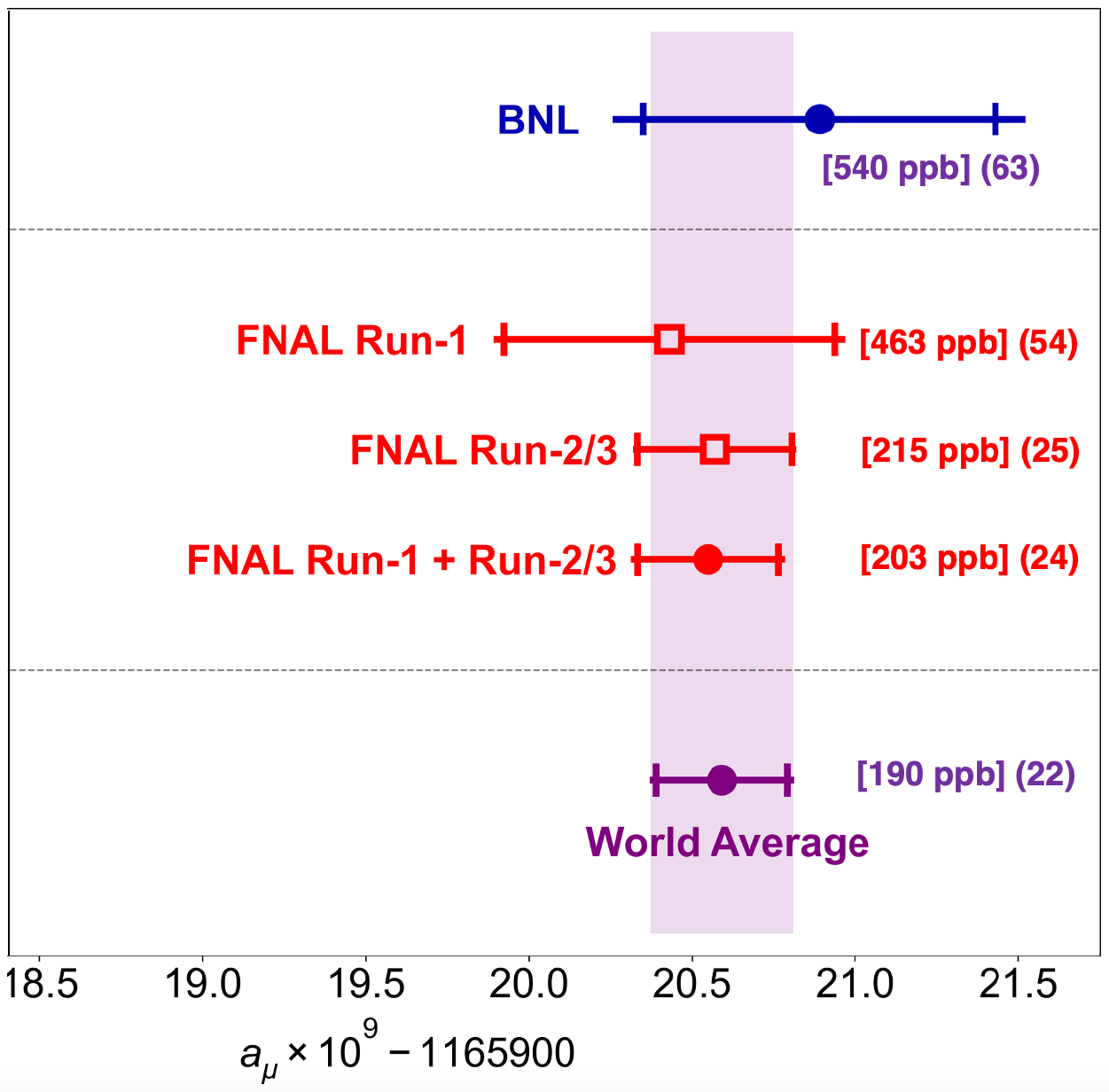}
    \caption{Experimental values of \amu\ from BNL E821~\cite{Muong-2:2006rrc}, our Run-1 result~\cite{Muong-2:2021ojo}, our new result~\cite{Muong-2:2023cdq} and the combined Fermilab one, and the new experimental average. In square (round) brackets the fractional (absolute) uncertainty. The absolute uncertainty in round brackets is in $10^{-11}$ units.}
    \label{result}
\end{figure}

As for Run-1,  Run-2/3 analysis was blinded to avoid unconscious biases. 
On the 24$^{th}$ July the Collaboration gathered in Liverpool for the Muon \gmtwo\ Physics week. The analyses were reviewed and there were not outstanding questions. We decided unanimously to proceed with the unblinding. Run-2/3 result has 215 ppb total error (a factor 2.2 better than Run-1) and is in excellent agreement with Run-1 result, so the two measurements were averaged resulting in 203 ppb total uncertainty, with a combined (BNL and FNAL) experimental average with 190 ppb precision, as shown in Fig~\ref{result}.
It is interesting to note that the current uncertainty on \amu\ ($22\times 10^{-11}$) is less than 1/6 of the Electroweak contribution to the muon \gmtwo\ ($153.6\times 10^{-11}$~\cite{Aoyama:2020ynm})\footnote{I thank my colleague H. Nguyen for pointing this out.}.

 \begin{figure}[htb]
\vspace{-3cm}
   \centering
    \includegraphics[width=1.\textwidth]{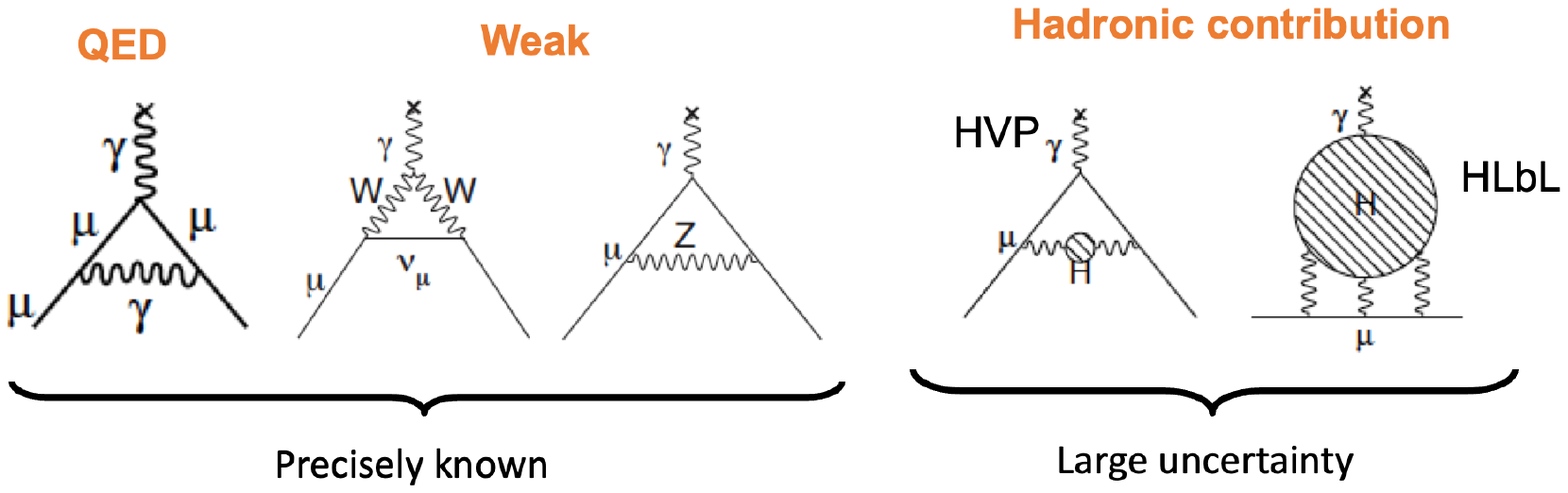}
\vspace{-3cm}
    \caption{SM contributions to the muon \gmtwo.}
    \label{fig_theory}
\end{figure}

\section{Comparison with the theory}
\label{theory}
\begin{figure}[htb]
   \centering
    \includegraphics[width=.48\textwidth]{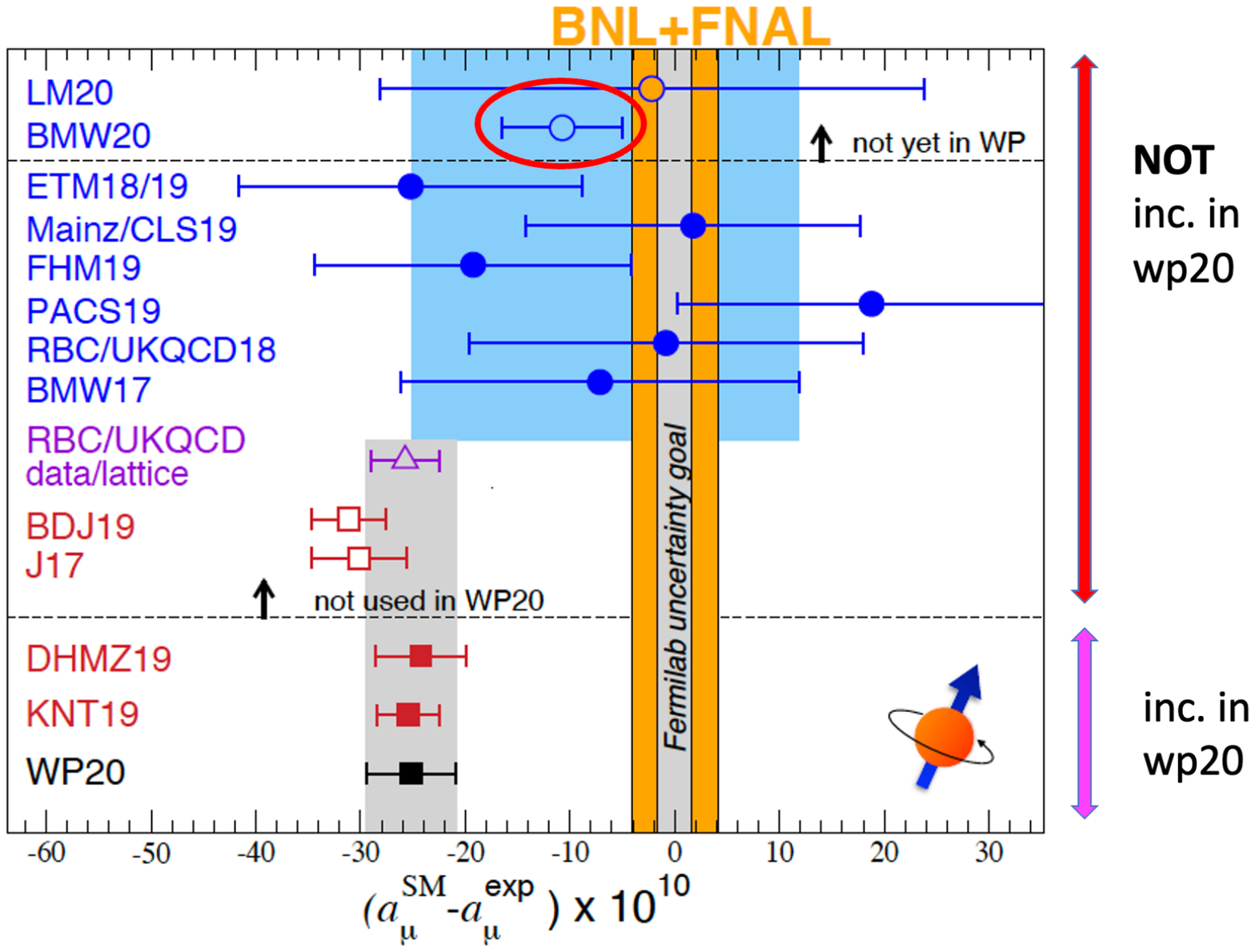}
   \includegraphics[width=.48\textwidth]{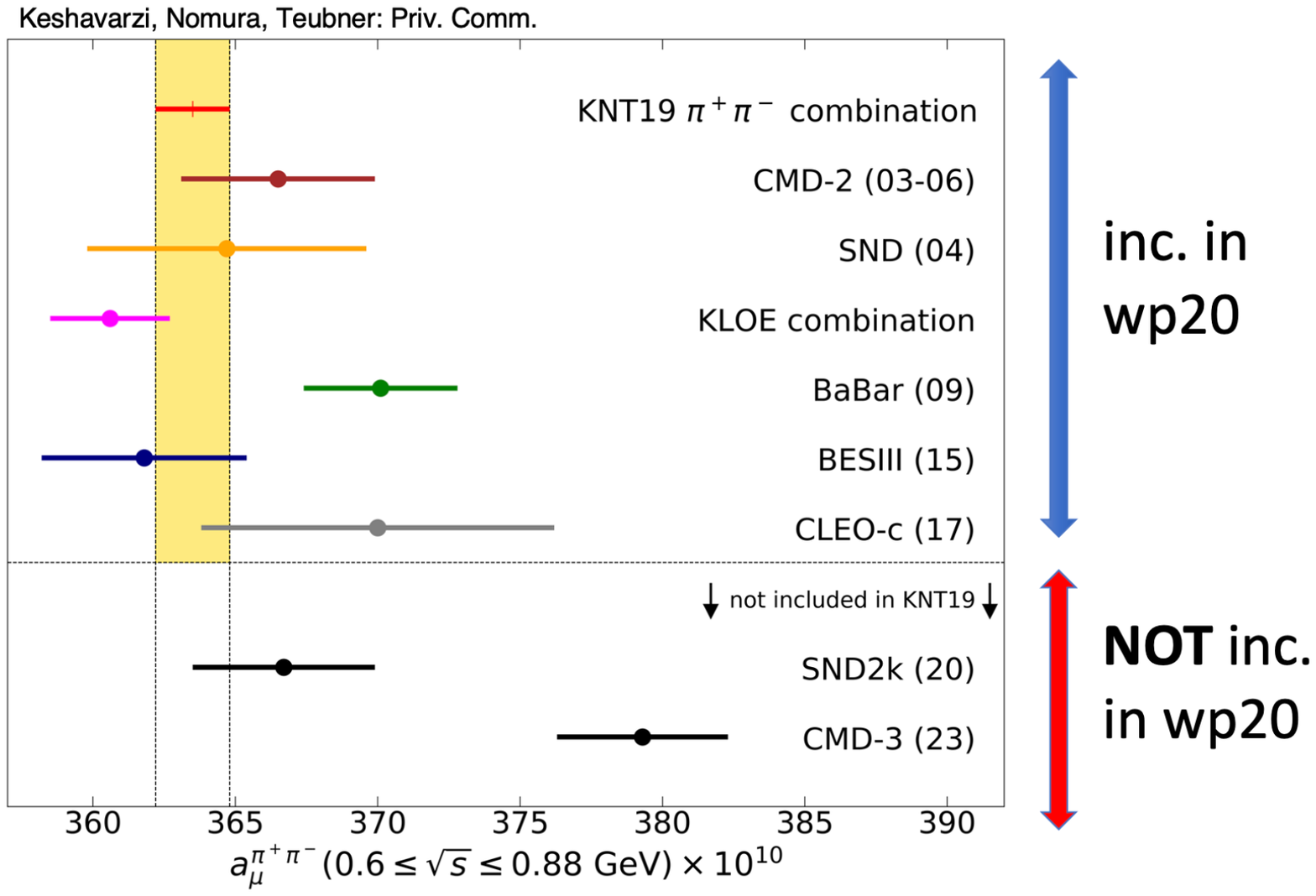}  
    \caption{Left: Comparison of theoretical predictions of \amu\ with the experimental value. Each data point represents a different evaluation of HVP. From Ref.~\cite{Colangelo:2022jxc}. Right: 2$\pi$ contribution to HVP from different $e^+e^-$ experiments.}
    \label{fig_theory2}
\end{figure}

Computing the muon \gmtwo\ is exceedingly complex. All sectors of the SM (electromagnetic, strong, and weak) provide contributions to the anomalous magnetic moment of leptons. 
In recent years, all aspects of the SM theory prediction for \amu\ have been scrutinized and refined with continued theoretical and computational efforts~\cite{Aoyama:2020ynm}. While the QED and electroweak contributions are widely considered non-controversial, the SM prediction of the muon \gmtwo\ is limited by our knowledge of the vacuum fluctuations involving strongly interacting particles, an effect called hadronic vacuum polarization (HVP), denoted also by \amuhlo (see Fig~\ref{fig_theory}).
In 2020 the Muon \gmtwo\ Theory Initiative, an international collaboration of more than 100 people, published an update of the SM prediction of the muon \gmtwo\ with accuracy of 0.37 ppm~\cite{Aoyama:2020ynm}. We refer to this prediction as wp20. 
The consensus prediction of 
 wp20~\cite{Aoyama:2020ynm} is based on the dispersive approach, for which experimental measurements of low-energy $e^+e^-\to hadrons$ cross sections serve as input. In 2021, the BMW collaboration published the first complete Lattice-QCD prediction of HVP with subpercent precision, that was closer to the experimental average and in $2.1\,\sigma$ tension with the prediction from the dispersive approach~\cite{Borsanyi:2020mff}, see Fig.~\ref{fig_theory2}, Left. In 2023, the CMD-3 experiment released a result on the $e^+e^-\to\pi^+\pi^-$ cross section that disagrees with all previous measurements used in the 2020 White Paper wp20, and for which the prediction of \amu\ is in less tension with the experimental value~\cite{CMD-3:2023alj}, see Fig.~\ref{fig_theory2}, Right.
If we compare the Run-2/3 result of \amu\ with wp20, it shows a significance of $5.1\,\sigma$. However current tensions and puzzles in the hadronic sector preclude  a firm comparison of the muon \gmtwo\ measurement with the theory.

\subsection{Outlook}
Analysis of Run-4/5/6 data, taken from  from December 2020 to July 2023, 
which is expected to be finalized by 2025 will bring an additional factor of two reduction on the statistical error which should allow us to reach  the total error of 140 ppb~\cite{Muong-2:2015xgu}. On a longer timescale a new measurement of the muon \gmtwo\ with a comparable accuracy of BNL is expected by the E34 experiment at J-PARC.
From the theoretical side  new results from lattice and analysis of $e^+e^-\to hadrons$ data should allow to clarify the prediction of the muon \gmtwo\, currently limited by puzzles in the hadronic sector. On a longer term new methods to compute HVP (like the MUonE experiment currently proposed at CERN~\cite{CarloniCalame:2015obs,Abbiendi:2016xup,pilato}), should provide independent inputs to the theoretical prediction.

\begin{figure}[htb]
   \centering
   \includegraphics[width=.8\textwidth]{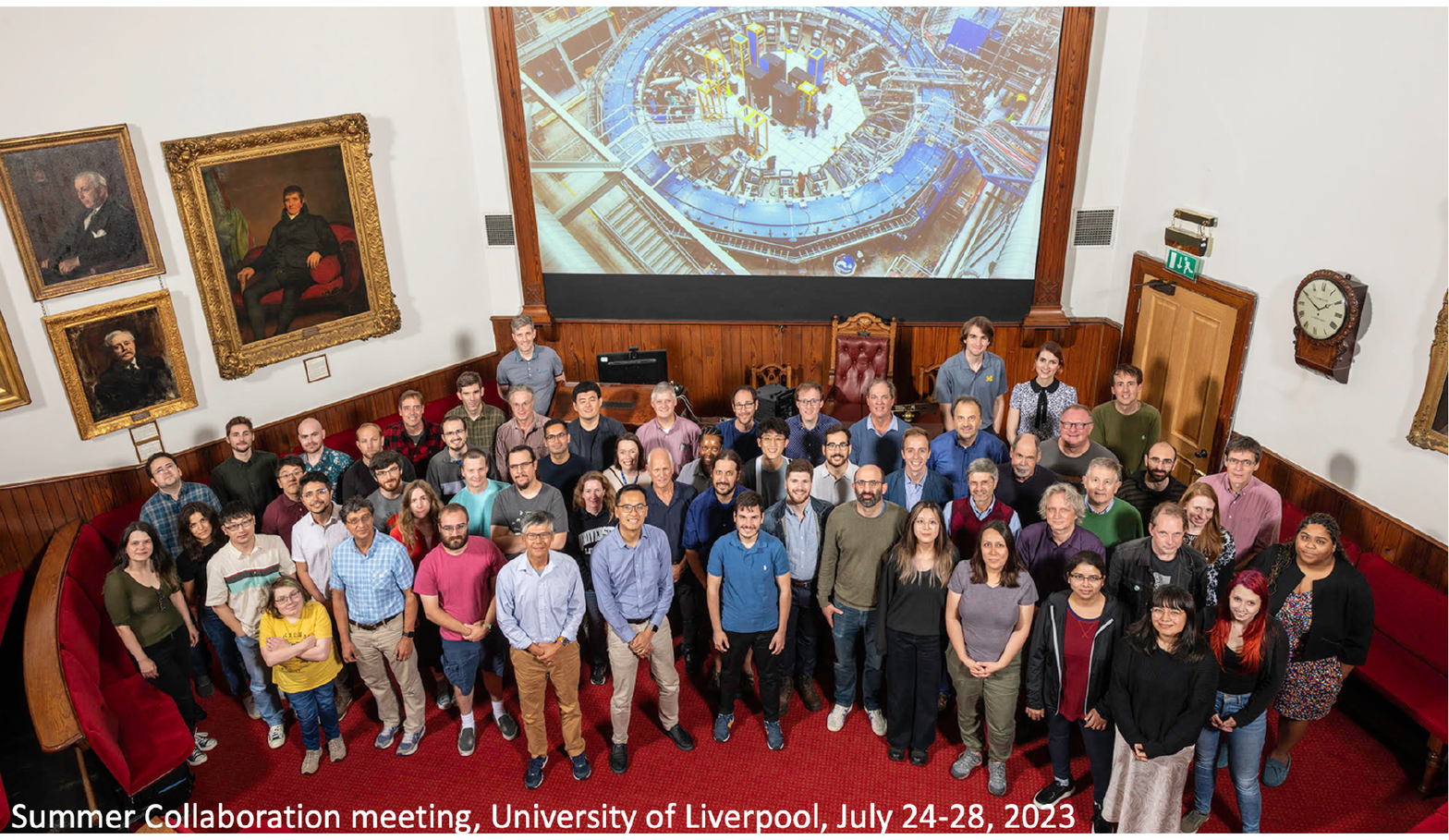}  
    \caption{Recent picture of the Muon \gmtwo\ Collaboration.}
    \label{collaboration}
\end{figure}

\section*{Acknowledgements}
I would like to thank the Muon \gmtwo\ collaborators who worked all together to measure one single beautiful number. Some of them are shown in Fig.~\ref{collaboration}. This work was supported by the Leverhulme Trust, LIP-2021-01.

\end{document}